\pdfoutput=1 
\documentclass[12pt]{article}
\usepackage[latin9]{inputenc}
\usepackage{graphicx}

\makeatletter

\providecommand{\tabularnewline}{\\}

\title{Continuous Post-Market Sequential Safety Surveillance with Minimum Events to Signal}

\date{ }

\author{ Martin Kulldorff$^{1,*}$ \and Ivair R. Silva$^{1,2}$
}

\makeatother

\graphicspath{%
    {converted_graphics/}
    {/}
}
\begin{document}
\maketitle


\begin{center}
$^{1}$ Department of Population Medicine, Harvard Medical School
and Harvard Pilgrim Health Care Institute, Boston, USA.\\
$^{2}$ Department of Statistics, Universidad Federal de Ouro Preto, Ouro Preto, Brazil.\\
 $^{*}$ martin\_kulldorff@hms.harvard.edu \\

\par\end{center}


\begin{abstract}
\noindent The CDC Vaccine Safety Datalink project has pioneered the
use of near real-time post-market vaccine safety surveillance for
the rapid detection of adverse events. Doing weekly analyses, continuous
sequential methods are used, allowing investigators to evaluate the
data near-continuously while still maintaining the correct overall
alpha level. With continuous sequential monitoring, the null hypothesis
may be rejected after only one or two adverse events are observed.
In this paper, we explore continuous sequential monitoring when 
we do not allow the null to be rejected until a minimum number of observed events have occurred.
We also evaluate continuous sequential analysis with a delayed start
until a certain sample size has been attained. Tables with exact critical
values, statistical power and the average times to signal are provided.
We show that, with the first option, it is possible to both increase
the power and reduce the expected time to signal, while keeping the
alpha level the same. The second option is only useful if the start of
the surveillance is delayed for logistical reasons, when there is
a group of data available at the first analysis, followed by continuous
or near-continuous monitoring thereafter. 
\end{abstract}
\noindent \textbf{Keywords:} Drug safety; Pharmacovigilance; Continuous
sequential analysis; Surveillance; Sequential probability ratio test.

\newpage{}

\section{Introduction}

Post-market drug and vaccine safety surveillance is important in order to detect
rare but serious adverse events not found during pre-licensure clinical
trials. Safety problems may go undetected either because an adverse reaction
is too rare to occur in sufficient numbers among the limited sample
size of a phase three clinical trial, or because the adverse reaction
only occur in a certain sub population that was excluded from the
trial, such as frail individuals.

In order to detect a safety problem as soon as possible, the CDC Vaccine
Safety Datalink project pioneered the use of near real-time safety
surveillance using automated weekly data feeds from electronic health
records\cite{Yih-Pediatrics2011,Davis-Pharmaceutics2013,McNeil-Vaccine2014}. In such surveillance, the goal is to detect serious adverse
reactions as early as possible without too many false signals. It
is then necessary to use sequential statistical analysis, which adjusts
for the multiple testing inherent in the many looks at the data. Using
the maximized sequential probability ratio test (MaxSPRT)\cite{ku-SA2011},
all new childhood vaccines and some adult vaccines are now monitored
in this fashion\cite{Yih-Pediatrics2011, Lieu-MC2007, Yih-Vaccine2009, Belongia-PIDJ2010, Klein-Pediatrics2010, Gee-Vaccine2011, Lee-AJPM2011, Tseng-Vaccine2013, Weintraub-NEJM2014, Daley-Vaccine2014}.
There is also interest in using sequential statistical methods
for post-market drug safety surveillance\cite{Brown-PDS2007, Avery-PDS2013, Fireman-PDS2012, Suling-Pharmaceutics2012, Gagne-Epidemiology2012, Gagne-PDS2014},
and the methods presented in this paper may also be used in either settings.

In contrast to group sequential analyses, continuous sequential methods
can signal after a single adverse event, if that event occurs sufficiently
early. In some settings, such as a phase 2 clinical trial, that may
be appropriate, but in post-market safety surveillance it is not. In post-market vaccine
surveillance, an ad-hoc rule that require at least two or three events
to signal has sometimes been used, but that leads to a conservative type 1 error (alpha
level). In this paper we provide exact critical values for continues
sequential analysis when a signal is required to have a certain minimum
number of adverse events. We also evaluate power and expected time to signal
for various alternative hypotheses. It is shown that it is possible
to simultaneously improve both of these by requiring at least 3 or
4 events to signal. Note that it is still necessary to start surveillance
as soon as the first few individuals are exposed, since they all could
have the adverse event.

For logistical reasons, there is sometimes a delay in the start of
post-marketing safety surveillance, so that the first analysis is
not conducted until a group of people have already been exposed to
the drug or vaccine. This is not a problem when using group sequential
methods, as the first group is then simply defined to correspond to
the start of surveillance. For continuous sequential surveillance,
a delayed start needs to be taken into account when calculating the
critical values. In this paper, we present exact critical values when
there is a delayed start in the sequential analysis. We also calculate
the power and time to signal for different relative risks.

In addition to ensuring that the sequential analysis maintains the
correct overall alpha level, it is important to consider the statistical
power to reject the null hypothesis; the average time until a signal
occurs when the null hypothesis is rejected; and the final sample
size when the null hypothesis is not rejected. For any fixed alpha,
there is a trade-off between these three metrics, and the trade-off
depends on the true relative risks. In clinical trials, where sequential
analyses are commonly used, statistical power and the final sample
size are usually the most important design criteria. The latter is
important because patient recruitment is costly. The time to signal
is usually the least important, as a slight delay in finding an adverse
event only affects the relatively small number of patients participating
in the clinical trial, but not the population-at-large. In post-market
safety surveillance, the trade-off is very different. Statistical
power is still very important, but once the surveillance system is
up and running, it is easy and cheap to prolong the length of the
study by a few extra months or years to achieve a final sample size
that provides the desired power. Instead, the second most critical metric
is the time to signal when the null is rejected. Since the product is already
in use by the population-at-large, most of which are not part of the surveillance 
system, a lot of people may be spared the adverse event if a safety problem 
can be detected a few weeks or months earlier. This means that for post-market
vaccine and drug safety surveillance, the final sample size when the
null is not rejected is the least important of the three metrics.

All calculations in this paper are exact, and none are based on simulations or asymptotic 
statistical theory. The numerical calculation of the exact critical values is a somewhat cumbersome
process. So that users do not have to do these calculations themselves,
we present tables with exact critical values for a wide range of parameters.
For other parameters, we have developed the open source R package 'Sequential',  
freely available at 'cran.r-project.org/web/packages/Sequential'.

\section{Continuous Sequential Analysis for Poisson Data}

Sequential analysis was first developed by Wald\cite{Wald-AMS1945,Wald-book1947},
who introduced the sequential probability ratio test (SPRT) for continuous
surveillance. The likelihood based SPRT proposed by Wald is very general
in that it can be used for many different probability distributions.
The SPRT is very sensitive to the definition of the alternative hypothesis
of a particular excess risk. For post-market safety surveillance,
a maximized sequential probability ratio test with a composite alternative
hypothesis has often been used instead. This is both a `generalized
sequential probability ratio test'\cite{Weiss-AMS1953} and `sequential
generalized likelihood ratio test'\cite{Lai-Ghosh1991,Siegmund-AS1980}.
In our setting, it is defined as follows, using the Poisson distribution
to model the number of adverse events seen\cite{ku-SA2011}.

Let $C_{t}$ be the random variable representing the number of adverse
events in a pre-defined risk window from $1$ to $W$ days after an
incident drug dispensing that was initiated during the time period
$[0,t]$. Let $c_{t}$ be the corresponding observed number of adverse
events. Note that time is defined in terms of the time of the drug
dispensing rather than the time of the adverse event, and that hence,
we actually do not know the value of $c_{t}$ until time $t+W$.

Under the null hypothesis ($H_{0}$), $C_{t}$ follows a Poisson distribution
with mean $\mu_{t}$, where $\mu_{t}$ is a known function reflecting
the population at risk. In our setting, $\mu_{t}$ reflects the number
of people who initiated their drug use during the time interval $[0,t]$
and a baseline risk for those individuals, adjusting for age, gender
and any other covariates of interest. Under the alternative hypothesis
($H_{A}$), the mean is instead $RR\mu_{t}$, where $RR$ is the increased
relative risk due to the drug/vaccine. Note that $C_{0}=c_{0}=\mu_{0}=0$.

For the Poisson model, the MaxSPRT likelihood ratio based test statistic
is 
\[
LR_{t}=\max_{H_{A}}\frac{P(C_{t}=c_{t}|H_{A})}{P(C_{t}=c_{t}|H_{0})}=\max_{RR>1}\frac{e^{-RR\mu_{t}}(RR\mu_{t})^{c_{t}}/c_{t}!}{e^{-\mu_{t}}\mu_{t}^{c_{t}}/c_{t}!}=\max_{RR>1}e^{(1-RR)\mu_{t}}(RR)^{c_{t}}
\]
The maximum likelihood estimate of $RR$ is $c_{t}/\mu_{t}$, so 
\[
LR_{t}=e^{\mu_{t}-c_{t}}(c_{t}/\mu_{t})^{c_{t}}
\]
Equivalently, when defined using the log likelihood ratio 
\[
LLR_{t}=ln(LR_{t})=\max_{RR>1}\left((1-RR)\mu_{t}+c_{t}ln(RR)\right)=(\mu_{t}-c_{t})+c_{t}ln(c_{t}/\mu_{t})
\]

This test statistic is sequentially monitored for all values of $t>0$,
until either $LLR_{t}\geq CV$, in which case the null hypothesis
is rejected, or until $\mu_{t}=T$, in which case the alternative
hypothesis is rejected. $T$ is a predefined upper limit on the length
of surveillance, defined in terms of the sample size, expressed as
the expected number of adverse events under the null hypothesis. It
is roughly equivalent to a certain number of exposed individuals,
but adjusted for covariates. Exact critical values (CV) are available
for the MaxSPRT\cite{ku-SA2011}, obtained through iterative numerical
calculations.

\section{Minimum Number of Events Required to Signal}

Continuous sequential probability ratio tests may signal at the time
of the first event, if that event appears sufficiently early. One
could add a requirement that there need to be a minimum of M events
before one can reject the null hypothesis. This still requires continuous
monitoring of the data from the very start, as $M$ events could appear
arbitrarily early. Hence, there is no logistical advantage of imposing 
this minimum number. The potential advantage is instead that
it may reduce the time to signal and/or increase the statistical power
of the study. Below, in Section 3.2, it is shown that both of these 
can be achieved simultaneously.

\subsection{Exact Critical Values}

\begin{table}
\begin{centering}
{\footnotesize{}}%
\begin{tabular}{c|ccccccc}
\hline 
 & \multicolumn{7}{c}{{\footnotesize{Minimum Number of Events Required to Reject the Null}}}\tabularnewline
{\footnotesize{T }} & {\footnotesize{M=1 }} & {\footnotesize{2 }} & {\footnotesize{3 }} & {\footnotesize{4 }} & {\footnotesize{6 }} & {\footnotesize{8 }} & {\footnotesize{10 }}\tabularnewline
\hline 
{\footnotesize{1 }} & {\footnotesize{2.853937 }} & {\footnotesize{2.366638 }} & {\footnotesize{1.774218 }} & {\footnotesize{.. }} & {\footnotesize{.. }} & {\footnotesize{.. }} & {\footnotesize{.. }}\tabularnewline
{\footnotesize{1.5 }} & {\footnotesize{2.964971 }} & {\footnotesize{2.576390 }} & {\footnotesize{2.150707 }} & {\footnotesize{1.683209 }} & {\footnotesize{.. }} & {\footnotesize{.. }} & {\footnotesize{.. }}\tabularnewline
{\footnotesize{2 }} & {\footnotesize{3.046977 }} & {\footnotesize{2.689354 }} & {\footnotesize{2.349679 }} & {\footnotesize{2.000158 }} & {\footnotesize{.. }} & {\footnotesize{.. }} & {\footnotesize{.. }}\tabularnewline
{\footnotesize{2.5 }} & {\footnotesize{3.110419 }} & {\footnotesize{2.777483 }} & {\footnotesize{2.474873 }} & {\footnotesize{2.187328 }} & {\footnotesize{.. }} & {\footnotesize{.. }} & {\footnotesize{.. }}\tabularnewline
{\footnotesize{3 }} & {\footnotesize{3.162106 }} & {\footnotesize{2.849327 }} & {\footnotesize{2.565320 }} & {\footnotesize{2.317139 }} & {\footnotesize{1.766485 }} & {\footnotesize{.. }} & {\footnotesize{.. }}\tabularnewline
{\footnotesize{4 }} & {\footnotesize{3.245004 }} & {\footnotesize{2.937410 }} & {\footnotesize{2.699182 }} & {\footnotesize{2.498892 }} & {\footnotesize{2.089473 }} & {\footnotesize{1.564636 }} & {\footnotesize{.. }}\tabularnewline
{\footnotesize{5 }} & {\footnotesize{3.297183 }} & {\footnotesize{3.012909 }} & {\footnotesize{2.803955 }} & {\footnotesize{2.623668 }} & {\footnotesize{2.267595 }} & {\footnotesize{1.936447 }} & {\footnotesize{.. }}\tabularnewline
{\footnotesize{6 }} & {\footnotesize{3.342729 }} & {\footnotesize{3.082099 }} & {\footnotesize{2.873904 }} & {\footnotesize{2.699350 }} & {\footnotesize{2.406810 }} & {\footnotesize{2.093835 }} & {\footnotesize{1.740551 }}\tabularnewline
{\footnotesize{8 }} & {\footnotesize{3.413782 }} & {\footnotesize{3.170062 }} & {\footnotesize{2.985560 }} & {\footnotesize{2.829259 }} & {\footnotesize{2.572627 }} & {\footnotesize{2.337771 }} & {\footnotesize{2.086032 }}\tabularnewline
{\footnotesize{10 }} & {\footnotesize{3.467952 }} & {\footnotesize{3.238009 }} & {\footnotesize{3.064248 }} & {\footnotesize{2.921561 }} & {\footnotesize{2.690586 }} & {\footnotesize{2.484834 }} & {\footnotesize{2.281441 }}\tabularnewline
{\footnotesize{12 }} & {\footnotesize{3.511749 }} & {\footnotesize{3.290551 }} & {\footnotesize{3.125253 }} & {\footnotesize{2.993106 }} & {\footnotesize{2.781435 }} & {\footnotesize{2.589388 }} & {\footnotesize{2.415402 }}\tabularnewline
{\footnotesize{15 }} & {\footnotesize{3.562591 }} & {\footnotesize{3.353265 }} & {\footnotesize{3.199953 }} & {\footnotesize{3.075613 }} & {\footnotesize{2.877939 }} & {\footnotesize{2.711996 }} & {\footnotesize{2.556634 }}\tabularnewline
{\footnotesize{20 }} & {\footnotesize{3.628123 }} & {\footnotesize{3.430141 }} & {\footnotesize{3.288216 }} & {\footnotesize{3.176370 }} & {\footnotesize{2.997792 }} & {\footnotesize{2.846858 }} & {\footnotesize{2.717137 }}\tabularnewline
{\footnotesize{25 }} & {\footnotesize{3.676320 }} & {\footnotesize{3.487961 }} & {\footnotesize{3.356677 }} & {\footnotesize{3.249634 }} & {\footnotesize{3.081051 }} & {\footnotesize{2.947270 }} & {\footnotesize{2.827711 }}\tabularnewline
{\footnotesize{30 }} & {\footnotesize{3.715764 }} & {\footnotesize{3.534150 }} & {\footnotesize{3.406715 }} & {\footnotesize{3.307135 }} & {\footnotesize{3.147801 }} & {\footnotesize{3.019639 }} & {\footnotesize{2.911222 }}\tabularnewline
{\footnotesize{40 }} & {\footnotesize{3.774663 }} & {\footnotesize{3.605056 }} & {\footnotesize{3.485960 }} & {\footnotesize{3.391974 }} & {\footnotesize{3.246619 }} & {\footnotesize{3.130495 }} & {\footnotesize{3.030735 }}\tabularnewline
{\footnotesize{50 }} & {\footnotesize{3.819903 }} & {\footnotesize{3.657142 }} & {\footnotesize{3.544826 }} & {\footnotesize{3.455521 }} & {\footnotesize{3.317955 }} & {\footnotesize{3.210428 }} & {\footnotesize{3.117553 }}\tabularnewline
{\footnotesize{60 }} & {\footnotesize{3.855755 }} & {\footnotesize{3.698885 }} & {\footnotesize{3.590567 }} & {\footnotesize{3.505220 }} & {\footnotesize{3.374194 }} & {\footnotesize{3.271486 }} & {\footnotesize{3.184196 }}\tabularnewline
{\footnotesize{80 }} & {\footnotesize{3.910853 }} & {\footnotesize{3.762474 }} & {\footnotesize{3.659939 }} & {\footnotesize{3.580900 }} & {\footnotesize{3.458087 }} & {\footnotesize{3.362888 }} & {\footnotesize{3.284030 }}\tabularnewline
{\footnotesize{100 }} & {\footnotesize{3.952321 }} & {\footnotesize{3.810141 }} & {\footnotesize{3.711993 }} & {\footnotesize{3.636508 }} & {\footnotesize{3.520081 }} & {\footnotesize{3.430065 }} & {\footnotesize{3.355794 }}\tabularnewline
{\footnotesize{120 }} & {\footnotesize{3.985577 }} & {\footnotesize{3.847748 }} & {\footnotesize{3.753329 }} & {\footnotesize{3.680584 }} & {\footnotesize{3.568679 }} & {\footnotesize{3.482966 }} & {\footnotesize{3.411235 }}\tabularnewline
{\footnotesize{150 }} & {\footnotesize{4.025338 }} & {\footnotesize{3.892715 }} & {\footnotesize{3.802412 }} & {\footnotesize{3.732386 }} & {\footnotesize{3.626150 }} & {\footnotesize{3.544308 }} & {\footnotesize{3.476655 }}\tabularnewline
{\footnotesize{200 }} & {\footnotesize{4.074828 }} & {\footnotesize{3.948930 }} & {\footnotesize{3.862762 }} & {\footnotesize{3.796835 }} & {\footnotesize{3.696511 }} & {\footnotesize{3.619825 }} & {\footnotesize{3.556799 }}\tabularnewline
{\footnotesize{250 }} & {\footnotesize{4.112234 }} & {\footnotesize{3.990901 }} & {\footnotesize{3.908065 }} & {\footnotesize{3.844847 }} & {\footnotesize{3.748757 }} & {\footnotesize{3.675703 }} & {\footnotesize{3.615513 }}\tabularnewline
{\footnotesize{300 }} & {\footnotesize{4.142134 }} & {\footnotesize{4.024153 }} & {\footnotesize{3.944135 }} & {\footnotesize{3.882710 }} & {\footnotesize{3.790143 }} & {\footnotesize{3.719452 }} & {\footnotesize{3.661830 }}\tabularnewline
{\footnotesize{400 }} & {\footnotesize{4.188031 }} & {\footnotesize{4.075297 }} & {\footnotesize{3.998950 }} & {\footnotesize{3.940563 }} & {\footnotesize{3.852658 }} & {\footnotesize{3.785930 }} & {\footnotesize{3.731524 }}\tabularnewline
{\footnotesize{500 }} & {\footnotesize{4.222632 }} & {\footnotesize{4.113692 }} & {\footnotesize{4.040021 }} & {\footnotesize{3.983778 }} & {\footnotesize{3.899239 }} & {\footnotesize{3.835265 }} & {\footnotesize{3.783126 }}\tabularnewline
{\footnotesize{600 }} & {\footnotesize{4.250310 }} & {\footnotesize{4.144317 }} & {\footnotesize{4.072638 }} & {\footnotesize{4.018090 }} & {\footnotesize{3.936175 }} & {\footnotesize{3.874183 }} & {\footnotesize{3.823908 }}\tabularnewline
{\footnotesize{800 }} & {\footnotesize{4.292829 }} & {\footnotesize{4.191167 }} & {\footnotesize{4.122559 }} & {\footnotesize{4.070466 }} & {\footnotesize{3.992272 }} & {\footnotesize{3.933364 }} & {\footnotesize{3.885600 }}\tabularnewline
{\footnotesize{1000 }} & {\footnotesize{4.324917 }} & {\footnotesize{4.226412 }} & {\footnotesize{4.160022 }} & {\footnotesize{4.109665 }} & {\footnotesize{4.034210 }} & {\footnotesize{3.977453 }} & {\footnotesize{3.931529 }}\tabularnewline
\hline 
\end{tabular}
\par\end{centering}{\footnotesize \par}

\centering{}{\footnotesize{\caption{{\footnotesize{Exact critical values for the Poisson based maximized
SPRT, when a minimum of M events is required before the null hypothesis
can be rejected. T is the upper limit on the sample size (length of
surveillance), expressed in terms of the expected number of events
under the null. The type 1 error is $\alpha=0.05$. When $T$ is small
and $M$ is large, no critical value will result in $\alpha\leq0.05$,
which is denoted by '..'\label{CVminimum}}}}
}}
\end{table}

In Table~\ref{CVminimum} we present the critical values for the
maximized SPRT when requiring a minimum number of events $M$ to signal.
When $M=1$, we get the standard maximized SPRT, whose previously calculated critical values\cite{ku-SA2011}
are included for comparison purposes.

The exact critical values are based on numerical calculations using
the R package 'Sequential'. The critical values were calculated in the same
manner as the exact critical values for the Poisson based MaxSPRT\cite{ku-SA2011},
with the modified requirement that the first possible time to signal
is at $M$ rather than $1$ event. In brief, first note that the time when the critical value is reached
and the null hypothesis is rejected can only happen at the time when
an event occurs. For any specified critical value $CV$ and maximum
sample size $T$, it is then possible to calculate alpha, the probability
of rejecting the null, using an iterative approach. Critical values
are then obtained through an iterative mathematical interpolation
process, until the desired precision is obtained. In 3 to 7 iterations,
the procedure converges to a precision of $0.00000001$. Note that
these numerical calculations only have to be done once for each alpha,
T and M. Hence, users do not need to do their own numerical calculations,
as long as they use one of the parameter combinations presented in
Table~\ref{CVminimum}.

The critical values are lower for higher values of $M$.
This is natural. Since we do not allow the null hypothesis to be rejected based on only
a small number of adverse events, it allows us to be more inclined to reject
the null later on when there are a larger number of events, while still maintaining
the correct overall alpha level. In essence, we are trading the ability to
reject the null with a very small number of events for the ability
to more easily reject the null when there are a medium or large number
of events. Note also that the critical values are higher for larger
values of the maximum sample size $T$. This is also natural, as there
is more multiple testing that needs to be adjusted for when $T$ is
large.

\subsection{Statistical Power and Expected Time to Signal}

\begin{table}
\begin{centering}
{\footnotesize{}}%
\begin{tabular}{cccccccccrrrrr}
\hline 
 &  &  & \multicolumn{5}{c}{{\footnotesize{Statistical Power}}} &  & \multicolumn{5}{c}{{\footnotesize{Average Time to Signal}}}\tabularnewline
{\footnotesize{T }} & {\footnotesize{M }} &  & {\footnotesize{RR=1.5 }} & {\footnotesize{2 }} & {\footnotesize{3 }} & {\footnotesize{4 }} & {\footnotesize{10 }} &  & {\footnotesize{RR=1.5 }} & {\footnotesize{2 }} & {\footnotesize{3 }} & {\footnotesize{4 }} & {\footnotesize{10}}\tabularnewline \hline
{\footnotesize{1 }} & {\footnotesize{1 }} & {\footnotesize{\vline}} & {\footnotesize{0.107 }} & {\footnotesize{0.185 }} & {\footnotesize{0.379 }} & {\footnotesize{0.573 }} & {\footnotesize{0.987 }} &  & {\footnotesize{0.30 }} & {\footnotesize{0.35}} & {\footnotesize{0.39 }} & {\footnotesize{0.39 }} & {\footnotesize{0.22 }}\tabularnewline
{\footnotesize{1 }} & {\footnotesize{3 }} & {\footnotesize{\vline}} & {\footnotesize{0.129 }} & {\footnotesize{0.234 }} & {\footnotesize{0.466 }} & {\footnotesize{0.665 }} & {\footnotesize{0.993 }} &  & {\footnotesize{0.59 }} & {\footnotesize{0.58}} & {\footnotesize{0.55 }} & {\footnotesize{0.51 }} & {\footnotesize{0.30 }}\tabularnewline
{\footnotesize{2 }} & {\footnotesize{1 }} & {\footnotesize{\vline}} & {\footnotesize{0.130 }} & {\footnotesize{0.255 }} & {\footnotesize{0.561 }} & {\footnotesize{0.799 }} & {\footnotesize{1.000 }} &  & {\footnotesize{0.63 }} & {\footnotesize{0.75}} & {\footnotesize{0.79 }} & {\footnotesize{0.73 }} & {\footnotesize{0.24 }}\tabularnewline
{\footnotesize{2 }} & {\footnotesize{3 }} & {\footnotesize{\vline}} & {\footnotesize{0.157 }} & {\footnotesize{0.315 }} & {\footnotesize{0.645 }} & {\footnotesize{0.857 }} & {\footnotesize{1.000 }} &  & {\footnotesize{0.92 }} & {\footnotesize{0.94}} & {\footnotesize{0.89 }} & {\footnotesize{0.78 }} & {\footnotesize{0.31 }}\tabularnewline
{\footnotesize{5 }} & {\footnotesize{1 }} & {\footnotesize{\vline}} & {\footnotesize{0.190 }} & {\footnotesize{0.447 }} & {\footnotesize{0.876 }} & {\footnotesize{0.987 }} & {\footnotesize{1.000 }} &  & {\footnotesize{1.82 }} & {\footnotesize{2.09}} & {\footnotesize{1.78 }} & {\footnotesize{1.22 }} & {\footnotesize{0.26 }}\tabularnewline
{\footnotesize{5 }} & {\footnotesize{3 }} & {\footnotesize{\vline}} & {\footnotesize{0.224 }} & {\footnotesize{0.507 }} & {\footnotesize{0.905 }} & {\footnotesize{0.991 }} & {\footnotesize{1.000 }} &  & {\footnotesize{2.10 }} & {\footnotesize{2.17}} & {\footnotesize{1.73 }} & {\footnotesize{1.17 }} & {\footnotesize{0.31 }}\tabularnewline
{\footnotesize{5 }} & {\footnotesize{6 }} & {\footnotesize{\vline}} & {\footnotesize{0.255 }} & {\footnotesize{0.559 }} & {\footnotesize{0.928 }} & {\footnotesize{0.994 }} & {\footnotesize{1.000 }} &  & {\footnotesize{2.71 }} & {\footnotesize{2.58}} & {\footnotesize{2.05 }} & {\footnotesize{1.54 }} & {\footnotesize{0.60 }}\tabularnewline
{\footnotesize{10 }} & {\footnotesize{1 }} & {\footnotesize{\vline}} & {\footnotesize{0.280 }} & {\footnotesize{0.685 }} & {\footnotesize{0.989 }} & {\footnotesize{1.000 }} & {\footnotesize{1.000 }} &  & {\footnotesize{4.02 }} & {\footnotesize{4.13}} & {\footnotesize{2.45 }} & {\footnotesize{1.35 }} & {\footnotesize{0.27 }}\tabularnewline
{\footnotesize{10 }} & {\footnotesize{3 }} & {\footnotesize{\vline}} & {\footnotesize{0.321 }} & {\footnotesize{0.733 }} & {\footnotesize{0.993 }} & {\footnotesize{1.000 }} & {\footnotesize{1.000 }} &  & {\footnotesize{4.25 }} & {\footnotesize{4.07}} & {\footnotesize{2.31 }} & {\footnotesize{1.30 }} & {\footnotesize{0.32 }}\tabularnewline
{\footnotesize{10 }} & {\footnotesize{6 }} & {\footnotesize{\vline}} & {\footnotesize{0.358 }} & {\footnotesize{0.770 }} & {\footnotesize{0.995 }} & {\footnotesize{1.000 }} & {\footnotesize{1.000 }} &  & {\footnotesize{4.71 }} & {\footnotesize{4.25}} & {\footnotesize{2.50 }} & {\footnotesize{1.61 }} & {\footnotesize{0.60 }}\tabularnewline
{\footnotesize{10 }} & {\footnotesize{10 }} & {\footnotesize{\vline}} & {\footnotesize{0.391 }} & {\footnotesize{0.803 }} & {\footnotesize{0.996 }} & {\footnotesize{1.000 }} & {\footnotesize{1.000 }} &  & {\footnotesize{5.67 }} & {\footnotesize{5.03}} & {\footnotesize{3.40 }} & {\footnotesize{2.50 }} & {\footnotesize{1.00 }}\tabularnewline
{\footnotesize{20 }} & {\footnotesize{1 }} & {\footnotesize{\vline}} & {\footnotesize{0.450 }} & {\footnotesize{0.921 }} & {\footnotesize{1.000 }} & {\footnotesize{1.000 }} & {\footnotesize{1.000 }} &  & {\footnotesize{8.68 }} & {\footnotesize{6.96}} & {\footnotesize{2.67 }} & {\footnotesize{1.41 }} & {\footnotesize{0.28 }}\tabularnewline
{\footnotesize{20 }} & {\footnotesize{3 }} & {\footnotesize{\vline}} & {\footnotesize{0.492 }} & {\footnotesize{0.936 }} & {\footnotesize{1.000 }} & {\footnotesize{1.000 }} & {\footnotesize{1.000 }} &  & {\footnotesize{8.65 }} & {\footnotesize{6.62}} & {\footnotesize{2.53 }} & {\footnotesize{1.37 }} & {\footnotesize{0.33 }}\tabularnewline
{\footnotesize{20 }} & {\footnotesize{6 }} & {\footnotesize{\vline}} & {\footnotesize{0.531 }} & {\footnotesize{0.948 }} & {\footnotesize{1.000 }} & {\footnotesize{1.000 }} & {\footnotesize{1.000 }} &  & {\footnotesize{8.92 }} & {\footnotesize{6.57}} & {\footnotesize{2.69 }} & {\footnotesize{1.65 }} & {\footnotesize{0.60 }}\tabularnewline
{\footnotesize{20 }} & {\footnotesize{10 }} & {\footnotesize{\vline}} & {\footnotesize{0.562 }} & {\footnotesize{0.957 }} & {\footnotesize{1.000 }} & {\footnotesize{1.000 }} & {\footnotesize{1.000 }} &  & {\footnotesize{9.47 }} & {\footnotesize{6.96}} & {\footnotesize{3.50 }} & {\footnotesize{2.51 }} & {\footnotesize{1.00 }}\tabularnewline
{\footnotesize{50 }} & {\footnotesize{1 }} & {\footnotesize{\vline}} & {\footnotesize{0.803 }} & {\footnotesize{1.000 }} & {\footnotesize{1.000 }} & {\footnotesize{1.000 }} & {\footnotesize{1.000 }} &  & {\footnotesize{20.45 }} & {\footnotesize{8.94}} & {\footnotesize{2.82 }} & {\footnotesize{1.48 }} & {\footnotesize{0.30 }}\tabularnewline
{\footnotesize{50 }} & {\footnotesize{3 }} & {\footnotesize{\vline}} & {\footnotesize{0.829 }} & {\footnotesize{1.000 }} & {\footnotesize{1.000 }} & {\footnotesize{1.000 }} & {\footnotesize{1.000 }} &  & {\footnotesize{19.82 }} & {\footnotesize{8.45}} & {\footnotesize{2.71 }} & {\footnotesize{1.45 }} & {\footnotesize{0.33 }}\tabularnewline
{\footnotesize{50 }} & {\footnotesize{6 }} & {\footnotesize{\vline}} & {\footnotesize{0.847 }} & {\footnotesize{1.000 }} & {\footnotesize{1.000 }} & {\footnotesize{1.000 }} & {\footnotesize{1.000 }} &  & {\footnotesize{19.41 }} & {\footnotesize{8.24}} & {\footnotesize{2.86 }} & {\footnotesize{1.71 }} & {\footnotesize{0.60 }}\tabularnewline
{\footnotesize{50 }} & {\footnotesize{10 }} & {\footnotesize{\vline}} & {\footnotesize{0.863 }} & {\footnotesize{1.000 }} & {\footnotesize{1.000 }} & {\footnotesize{1.000 }} & {\footnotesize{1.000 }} &  & {\footnotesize{19.35 }} & {\footnotesize{8.46}} & {\footnotesize{3.59 }} & {\footnotesize{2.52 }} & {\footnotesize{1.00 }}\tabularnewline
{\footnotesize{100 }} & {\footnotesize{1 }} & {\footnotesize{\vline}} & {\footnotesize{0.978 }} & {\footnotesize{1.000 }} & {\footnotesize{1.000 }} & {\footnotesize{1.000 }} & {\footnotesize{1.000 }} &  & {\footnotesize{29.93 }} & {\footnotesize{9.30}} & {\footnotesize{2.92 }} & {\footnotesize{1.53 }} & {\footnotesize{0.31 }}\tabularnewline
{\footnotesize{100 }} & {\footnotesize{3 }} & {\footnotesize{\vline}} & {\footnotesize{0.982 }} & {\footnotesize{1.000 }} & {\footnotesize{1.000 }} & {\footnotesize{1.000 }} & {\footnotesize{1.000 }} &  & {\footnotesize{28.52 }} & {\footnotesize{8.87}} & {\footnotesize{2.82 }} & {\footnotesize{1.51 }} & {\footnotesize{0.34 }}\tabularnewline
{\footnotesize{100 }} & {\footnotesize{6 }} & {\footnotesize{\vline}} & {\footnotesize{0.985 }} & {\footnotesize{1.000 }} & {\footnotesize{1.000 }} & {\footnotesize{1.000 }} & {\footnotesize{1.000 }} &  & {\footnotesize{27.58 }} & {\footnotesize{8.71}} & {\footnotesize{2.97 }} & {\footnotesize{1.75 }} & {\footnotesize{0.60 }}\tabularnewline
{\footnotesize{100 }} & {\footnotesize{10 }} & {\footnotesize{\vline}} & {\footnotesize{0.987 }} & {\footnotesize{1.000 }} & {\footnotesize{1.000 }} & {\footnotesize{1.000 }} & {\footnotesize{1.000 }} &  & {\footnotesize{27.04 }} & {\footnotesize{8.93}} & {\footnotesize{3.65 }} & {\footnotesize{2.53 }} & {\footnotesize{1.00 }}\tabularnewline
{\footnotesize{200 }} & {\footnotesize{1 }} & {\footnotesize{\vline}} & {\footnotesize{1.000 }} & {\footnotesize{1.000 }} & {\footnotesize{1.000 }} & {\footnotesize{1.000 }} & {\footnotesize{1.000 }} &  & {\footnotesize{33.00 }} & {\footnotesize{9.62}} & {\footnotesize{3.01 }} & {\footnotesize{1.58 }} & {\footnotesize{0.32 }}\tabularnewline
{\footnotesize{200 }} & {\footnotesize{3 }} & {\footnotesize{\vline}} & {\footnotesize{1.000 }} & {\footnotesize{1.000 }} & {\footnotesize{1.000 }} & {\footnotesize{1.000 }} & {\footnotesize{1.000 }} &  & {\footnotesize{31.47 }} & {\footnotesize{9.25}} & {\footnotesize{2.93 }} & {\footnotesize{1.56 }} & {\footnotesize{0.35 }}\tabularnewline
{\footnotesize{200 }} & {\footnotesize{6 }} & {\footnotesize{\vline}} & {\footnotesize{1.000 }} & {\footnotesize{1.000 }} & {\footnotesize{1.000 }} & {\footnotesize{1.000 }} & {\footnotesize{1.000 }} &  & {\footnotesize{30.47 }} & {\footnotesize{9.11}} & {\footnotesize{3.07 }} & {\footnotesize{1.78 }} & {\footnotesize{0.60 }}\tabularnewline
{\footnotesize{200 }} & {\footnotesize{10 }} & {\footnotesize{\vline}} & {\footnotesize{1.000 }} & {\footnotesize{1.000 }} & {\footnotesize{1.000 }} & {\footnotesize{1.000 }} & {\footnotesize{1.000 }} &  & {\footnotesize{29.88 }} & {\footnotesize{9.33}} & {\footnotesize{3.71 }} & {\footnotesize{2.54 }} & {\footnotesize{1.00 }}\tabularnewline
\hline 
\end{tabular}
\par\end{centering}{\footnotesize \par}

\centering{}{\footnotesize{\caption{{\footnotesize{Statistical power and average time to signal, when
the null hypothesis is rejected, for the Poisson based maximized SPRT
when a minimum of M events is required before the null hypothesis
can be rejected. T is the upper limit on the sample size (length of
surveillance), expressed in terms of the expected number of events
under the null. The type 1 error is $\alpha=0.05$. \label{PTminimum} }}}
}}
\end{table}

In Table~\ref{PTminimum} we present statistical power and average
time to signal for different values of $M$, the minimum number of
events needed to signal. These are exact calculations, done for different
relative risks and for different upper limits $T$ on the length of
surveillance. When $T$ increases, power increases, since the maximum
sample size increases. For fixed $T$, the power always increases
with increasing $M$. This is natural, since power increases by default
when there are fewer looks at the data, as there is less multiple
testing to adjust for. The average time to signal may either increase
or decrease with increasing values of $M$. For example, with $T=20$
and a true $RR=2$, the average time of signal is $6.96$, $6.62$,
$6.57$ and $6.96$ for $M=1,3,6$ and $10$, respectively. For the
same parameters, the statistical power is $0.921$, $0.936$, $0.948$
and $0.957$ respectively. Hence, when the true $RR=2$ and when $T=20$,
both power and the average time to signal is better if we use $M=3$
rather than $M=1$. The same is true for $M=6$ versus $M=3$, but
not for $M=10$ versus $M=6$.

\begin{figure}[tbp] 
  \centering
  \includegraphics[bb=0 0 1280 923,width=5.67in,height=4.09in,keepaspectratio]{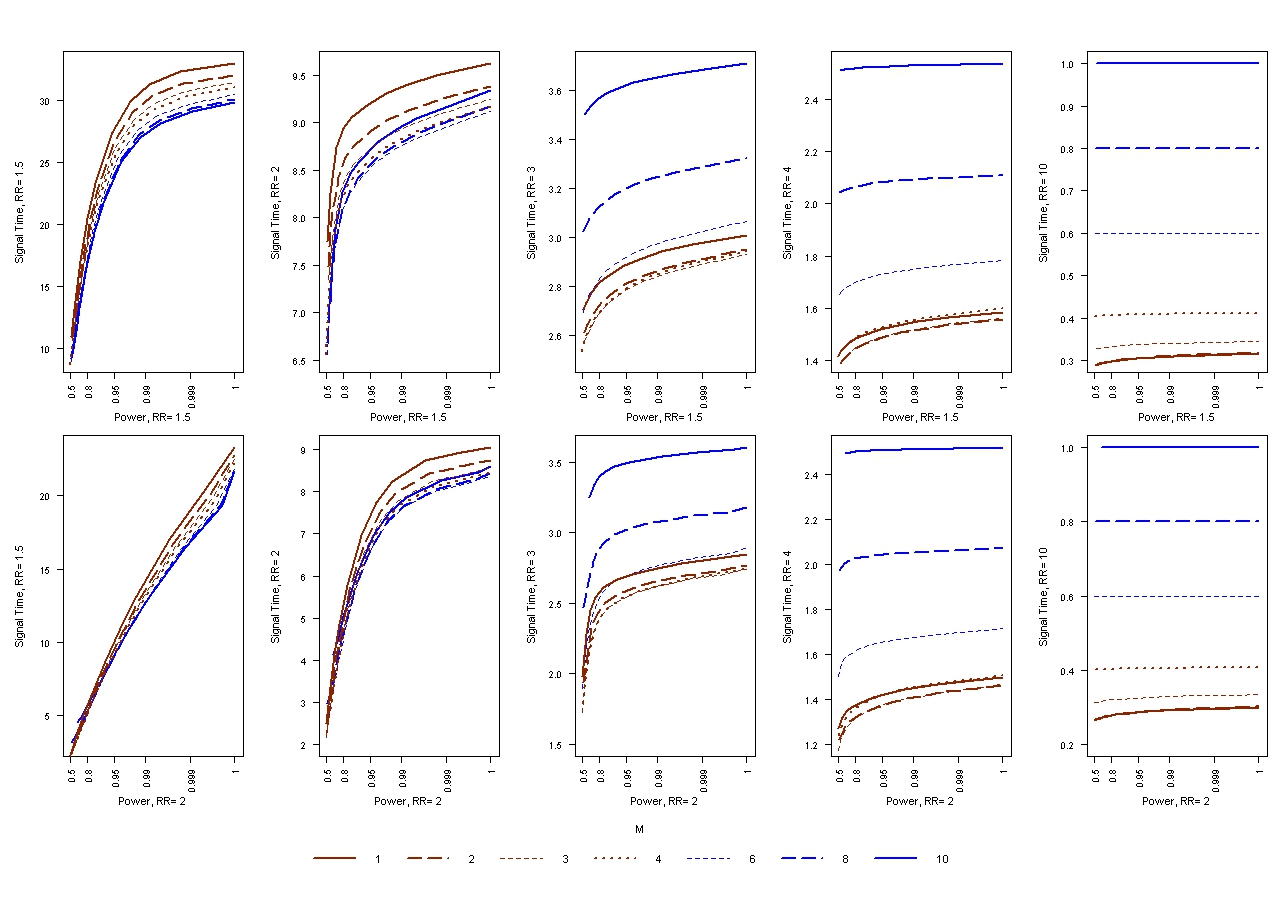}
  \caption{The average time to signal, as a function of statistical power, for
the Poisson based MaxSPRT when a minimum of M events is required
before the null hypothesis can be rejected. The type 1 error is $\alpha=0.05$.
\label{minimum}}
\end{figure}

The trade-off between statistical power and average time to signal
is not easily deciphered from Table~\ref{PTminimum}, and it is hence
hard to judge which value of $M$ is best. Since $T$, the upper limit
on the length of surveillance, is the least important metric, let's
ignore that for the moment, and see what happens to the average time
to signal if we keep both the alpha level and the power fixed. That
will make it easier to find a good choice for $M$, which will depend
on the true relative risk. Figure~\ref{minimum} shows the average
time to signal as a function of statistical power, for different values
of $M$. The lower curves are better, since the expected time to signal
is shorter. Suppose we design the sequential analysis to have 95 percent
power to detect a relative risk of 1.5. We can then look at the left
side of Figure~\ref{minimum} to see the average time to signal for
different true relative risks. We see that for a true relative risk
of 1.5, time to signal is shortest for $M=10$. On the other hand,
for a true relative risk of 2, it is shortest for $M=6$, for a true
relative risk of 3, it is shortest for $M=3$ and for a true relative
risk of 4, it is shortest for $M=2$. On the right side of Figure~\ref{minimum},
we show the expected time to signal when the surveillance has been
designed to attain a certain power for a relative risk of 2. The results
are similar.

When the true relative risk is higher, it is a more serious safety
problem, and hence, it is more important to detect it earlier. So,
while there is no single value of $M$ that is best overall, anywhere
in the 3 to 6 range may be a reasonable choice for $M$. The cost
of this reduced time to signal when the null is rejected is a slight
delay until the surveillance ends when the null is not rejected.

\section{Delayed Start of Surveillance}

For logistical or other reasons, it is not always possible to start
post-marketing safety surveillance at the time that the first vaccine
or drug is given. If the delay is short, one could ignore this and
pretend that the sequential analyses started with the first exposed
person. One could do this either by starting to calculate the test
statistic at time $D$ or by calculating it retroactively for all
times before $D$. The former will be conservative, not maintaining
the correct alpha level. The latter will maintain the correct alpha
level, but, some signals will be unnecessarily delayed without a compensatory
improvement in any of the other metrics. A better solution is to use
critical values that take the delayed start of surveillance into account.

\subsection{Exact Critical Values}

\begin{table}
\begin{centering}
{\footnotesize{}}%
\begin{tabular}{c|cccccccc}
\hline 
 & \multicolumn{8}{c}{{\footnotesize{D}}}\tabularnewline
{\footnotesize{T }} & {\footnotesize{0 }} & {\footnotesize{1 }} & {\footnotesize{2 }} & {\footnotesize{3 }} & {\footnotesize{4 }} & {\footnotesize{6 }} & {\footnotesize{8 }} & {\footnotesize{10 }}\tabularnewline
\hline 
{\footnotesize{1.5 }} & {\footnotesize{2.964971 }} & {\footnotesize{1.683208 }} & {\footnotesize{.. }} & {\footnotesize{.. }} & {\footnotesize{.. }} & {\footnotesize{.. }} & {\footnotesize{.. }} & {\footnotesize{.. }}\tabularnewline
{\footnotesize{2 }} & {\footnotesize{3.046977 }} & {\footnotesize{2.000158 }} & {\footnotesize{.. }} & {\footnotesize{.. }} & {\footnotesize{.. }} & {\footnotesize{.. }} & {\footnotesize{.. }} & {\footnotesize{.. }}\tabularnewline
{\footnotesize{2.5 }} & {\footnotesize{3.110419 }} & {\footnotesize{2.187328 }} & {\footnotesize{1.600544 }} & {\footnotesize{.. }} & {\footnotesize{.. }} & {\footnotesize{.. }} & {\footnotesize{.. }} & {\footnotesize{.. }}\tabularnewline
{\footnotesize{3 }} & {\footnotesize{3.162106 }} & {\footnotesize{2.317139 }} & {\footnotesize{1.766484 }} & {\footnotesize{.. }} & {\footnotesize{.. }} & {\footnotesize{.. }} & {\footnotesize{.. }} & {\footnotesize{.. }}\tabularnewline
{\footnotesize{4 }} & {\footnotesize{3.245004 }} & {\footnotesize{2.498892 }} & {\footnotesize{2.089473 }} & {\footnotesize{1.842319 }} & {\footnotesize{.. }} & {\footnotesize{.. }} & {\footnotesize{.. }} & {\footnotesize{.. }}\tabularnewline
{\footnotesize{5 }} & {\footnotesize{3.297183 }} & \textit{\footnotesize{2.545178}}{\footnotesize{ }} & {\footnotesize{2.267595 }} & {\footnotesize{1.936447 }} & {\footnotesize{1.611553 }} & {\footnotesize{.. }} & {\footnotesize{.. }} & {\footnotesize{.. }}\tabularnewline
{\footnotesize{6 }} & {\footnotesize{3.342729 }} & {\footnotesize{2.546307 }} & {\footnotesize{2.406809 }} & {\footnotesize{2.093835 }} & {\footnotesize{1.921859 }} & {\footnotesize{.. }} & {\footnotesize{.. }} & {\footnotesize{.. }}\tabularnewline
{\footnotesize{8 }} & {\footnotesize{3.413782 }} & {\footnotesize{2.694074 }} & {\footnotesize{2.572627 }} & {\footnotesize{2.337771 }} & {\footnotesize{2.211199 }} & {\footnotesize{1.829011 }} & {\footnotesize{.. }} & {\footnotesize{.. }}\tabularnewline
{\footnotesize{10 }} & {\footnotesize{3.467952 }} & {\footnotesize{2.799333 }} & \textit{\footnotesize{2.591675}}{\footnotesize{ }} & {\footnotesize{2.484834 }} & \textit{\footnotesize{2.298373}}{\footnotesize{ }} & {\footnotesize{2.087405 }} & \textit{\footnotesize{1.834622}}{\footnotesize{ }} & {\footnotesize{.. }}\tabularnewline
{\footnotesize{12 }} & {\footnotesize{3.511749 }} & {\footnotesize{2.880721 }} & {\footnotesize{2.683713 }} & {\footnotesize{2.589388 }} & {\footnotesize{2.415402 }} & {\footnotesize{2.254018 }} & {\footnotesize{1.965660 }} & {\footnotesize{1.755455 }}\tabularnewline
{\footnotesize{15 }} & {\footnotesize{3.562591 }} & {\footnotesize{2.970411 }} & {\footnotesize{2.794546 }} & {\footnotesize{2.711996 }} & {\footnotesize{2.556634 }} & {\footnotesize{2.347591 }} & {\footnotesize{2.203782 }} & \textit{\footnotesize{2.020681}}{\footnotesize{ }}\tabularnewline
{\footnotesize{20 }} & {\footnotesize{3.628123 }} & {\footnotesize{3.082511 }} & {\footnotesize{2.918988 }} & \textit{\footnotesize{2.846635}}{\footnotesize{ }} & {\footnotesize{2.717137 }} & {\footnotesize{2.542045 }} & {\footnotesize{2.425671 }} & {\footnotesize{2.260811 }}\tabularnewline
{\footnotesize{25 }} & {\footnotesize{3.676320 }} & {\footnotesize{3.159490 }} & {\footnotesize{3.011001 }} & {\footnotesize{2.886783 }} & {\footnotesize{2.827711 }} & {\footnotesize{2.668487 }} & {\footnotesize{2.527763 }} & {\footnotesize{2.432668 }}\tabularnewline
{\footnotesize{30 }} & {\footnotesize{3.715764 }} & {\footnotesize{3.223171 }} & {\footnotesize{3.080629 }} & {\footnotesize{2.963485 }} & {\footnotesize{2.911222 }} & {\footnotesize{2.765594 }} & {\footnotesize{2.634068 }} & {\footnotesize{2.553373 }}\tabularnewline
{\footnotesize{40 }} & {\footnotesize{3.774663 }} & {\footnotesize{3.313966 }} & {\footnotesize{3.186878 }} & {\footnotesize{3.078748 }} & {\footnotesize{3.030735 }} & {\footnotesize{2.903286 }} & {\footnotesize{2.789967 }} & {\footnotesize{2.684730 }}\tabularnewline
{\footnotesize{50 }} & {\footnotesize{3.819903 }} & {\footnotesize{3.381606 }} & {\footnotesize{3.261665 }} & {\footnotesize{3.162197 }} & {\footnotesize{3.117553 }} & {\footnotesize{2.999580 }} & {\footnotesize{2.897811 }} & {\footnotesize{2.802863 }}\tabularnewline
{\footnotesize{60 }} & {\footnotesize{3.855755 }} & {\footnotesize{3.434748 }} & {\footnotesize{3.320749 }} & {\footnotesize{3.226113 }} & \textit{\footnotesize{3.162908}}{\footnotesize{ }} & \textit{\footnotesize{3.051470}}{\footnotesize{ }} & {\footnotesize{2.978063 }} & {\footnotesize{2.890933 }}\tabularnewline
{\footnotesize{80 }} & {\footnotesize{3.910853 }} & {\footnotesize{3.515052 }} & {\footnotesize{3.407923 }} & {\footnotesize{3.321868 }} & {\footnotesize{3.247872 }} & {\footnotesize{3.151820 }} & \textit{\footnotesize{3.090356}}{\footnotesize{ }} & {\footnotesize{3.019184 }}\tabularnewline
{\footnotesize{100 }} & {\footnotesize{3.952321 }} & {\footnotesize{3.574091 }} & {\footnotesize{3.472610 }} & {\footnotesize{3.391377 }} & {\footnotesize{3.321971 }} & {\footnotesize{3.232345 }} & {\footnotesize{3.155596 }} & {\footnotesize{3.109251 }}\tabularnewline
{\footnotesize{120 }} & {\footnotesize{3.985577 }} & {\footnotesize{3.620223 }} & {\footnotesize{3.523446 }} & {\footnotesize{3.445695 }} & {\footnotesize{3.379278 }} & {\footnotesize{3.294843 }} & {\footnotesize{3.222053 }} & {\footnotesize{3.177847 }}\tabularnewline
{\footnotesize{150 }} & {\footnotesize{4.025338 }} & {\footnotesize{3.675035 }} & {\footnotesize{3.583195 }} & {\footnotesize{3.509028 }} & {\footnotesize{3.446674 }} & {\footnotesize{3.367227 }} & {\footnotesize{3.298671 }} & {\footnotesize{3.238461 }}\tabularnewline
{\footnotesize{200 }} & {\footnotesize{4.074828 }} & {\footnotesize{3.742843 }} & {\footnotesize{3.655984 }} & {\footnotesize{3.587079 }} & {\footnotesize{3.528662 }} & {\footnotesize{3.454679 }} & {\footnotesize{3.391821 }} & {\footnotesize{3.336012 }}\tabularnewline
{\footnotesize{250 }} & {\footnotesize{4.112234 }} & {\footnotesize{3.792978 }} & {\footnotesize{3.710128 }} & {\footnotesize{3.644349 }} & {\footnotesize{3.588871 }} & {\footnotesize{3.518954 }} & {\footnotesize{3.459256 }} & {\footnotesize{3.406929 }}\tabularnewline
{\footnotesize{300 }} & {\footnotesize{4.142134 }} & {\footnotesize{3.832686 }} & {\footnotesize{3.752749 }} & {\footnotesize{3.689355 }} & {\footnotesize{3.636272 }} & {\footnotesize{3.568952 }} & {\footnotesize{3.512138 }} & {\footnotesize{3.462111 }}\tabularnewline
{\footnotesize{400 }} & {\footnotesize{4.188031 }} & {\footnotesize{3.893093 }} & {\footnotesize{3.785930 }} & {\footnotesize{3.757574 }} & {\footnotesize{3.707431 }} & {\footnotesize{3.644405 }} & {\footnotesize{3.591092 }} & {\footnotesize{3.544518 }}\tabularnewline
{\footnotesize{500 }} & {\footnotesize{4.222632 }} & {\footnotesize{3.938105 }} & {\footnotesize{3.835264 }} & {\footnotesize{3.808087 }} & {\footnotesize{3.760123 }} & {\footnotesize{3.700032 }} & {\footnotesize{3.649189 }} & {\footnotesize{3.605012 }}\tabularnewline
{\footnotesize{600 }} & {\footnotesize{4.250310 }} & {\footnotesize{3.973710 }} & {\footnotesize{3.874183 }} & {\footnotesize{3.847892 }} & {\footnotesize{3.801678 }} & {\footnotesize{3.743656 }} & {\footnotesize{3.694832 }} & {\footnotesize{3.652326 }}\tabularnewline
{\footnotesize{800 }} & {\footnotesize{4.292829 }} & {\footnotesize{4.028089 }} & {\footnotesize{3.933363 }} & \textit{\footnotesize{3.887512}}{\footnotesize{ }} & {\footnotesize{3.864597 }} & {\footnotesize{3.809685 }} & {\footnotesize{3.763627 }} & {\footnotesize{3.723608 }}\tabularnewline
{\footnotesize{1000 }} & {\footnotesize{4.324917 }} & \textit{\footnotesize{4.047191}}{\footnotesize{ }} & {\footnotesize{3.977453 }} & {\footnotesize{3.931529 }} & {\footnotesize{3.911308 }} & {\footnotesize{3.858669 }} & \textit{\footnotesize{3.814122}}{\footnotesize{ }} & {\footnotesize{3.776275 }}\tabularnewline
\hline 
\end{tabular}
\par\end{centering}{\footnotesize \par}

\centering{}{\footnotesize{\caption{{\footnotesize{Exact critical values for the Poisson based maximized
SPRT, when surveillance does not start until the sample size is large
enough to generate D expected events under the null hypothesis. $T>D$
is the upper limit on the sample size. The minimum number of events
needed to reject is set to $M=1$. The type 1 error is $\alpha=0.05$.
For some values of $T$ and $D$, the critical values are conservative
with $\alpha<0.05$. These are denoted in italics.\label{CVdelay} }}}
}}
\end{table}

In Table~\ref{CVdelay} we present exact critical values for the
maximized SPRT when surveillance does not start until the expected
number of events under the null hypothesis is $D$, without any requirement
on having a minimum umber of events to signal. When $D=0$, we get
the standard maximized SPRT, whose critical values\cite{ku-SA2011}
are included for comparison purposes. Note that the critical values
are lower for higher values of $D$. Since surveillance is not performed
until the sample size have reached $D$ expected counts under the
null, one can afford to use a lower critical value for the remaining
time while still maintaining the same overall alpha level. As before,
the critical values are higher for larger values of $T$. When $D>T$,
the surveillance would not start until after the end of surveillance,
so those entries are blank in Table~\ref{CVdelay}. When $D=T$,
there is only one non-sequential analysis performed, so there are
no critical values for a sequential test procedure. Hence, they are
also left blank in the Table.

\begin{table}
\begin{centering}
{\footnotesize{}}%
\begin{tabular}{ccc|ccccccc}
\hline 
{\footnotesize{T }} & {\footnotesize{D }} & {\footnotesize{M}} & {\footnotesize{$CV_{cons}$ }} & {\footnotesize{$\alpha_{cons}$ }} & {\footnotesize{$CV_{lib}$ }} & {\footnotesize{$\alpha_{lib}$ }} &  &  & \tabularnewline
\hline 
{\footnotesize{5 }} & {\footnotesize{1 }} & {\footnotesize{1,4}} & {\footnotesize{2.545178 }} & {\footnotesize{0.04587 }} & {\footnotesize{2.545177 }} & {\footnotesize{0.05323 }} &  &  & \tabularnewline
{\footnotesize{10 }} & {\footnotesize{2 }} & {\footnotesize{1,4}} & {\footnotesize{2.591675 }} & {\footnotesize{0.04998 }} & {\footnotesize{2.591674 }} & {\footnotesize{0.05478 }} &  &  & \tabularnewline
{\footnotesize{10 }} & {\footnotesize{4 }} & {\footnotesize{1,4}} & {\footnotesize{2.298373 }} & {\footnotesize{0.04924 }} & {\footnotesize{2.298372 }} & {\footnotesize{0.05379 }} &  &  & \tabularnewline
{\footnotesize{10 }} & {\footnotesize{8 }} & {\footnotesize{1,4}} & {\footnotesize{1.834622 }} & {\footnotesize{0.04373 }} & {\footnotesize{1.834621 }} & {\footnotesize{0.05001 }} &  &  & \tabularnewline
{\footnotesize{15 }} & {\footnotesize{10 }} & {\footnotesize{1,4}} & {\footnotesize{2.020681 }} & {\footnotesize{0.04755 }} & {\footnotesize{2.020680 }} & {\footnotesize{0.05124 }} &  &  & \tabularnewline
{\footnotesize{20 }} & {\footnotesize{3 }} & {\footnotesize{1,4}} & {\footnotesize{2.846635 }} & {\footnotesize{0.04712 }} & {\footnotesize{2.846634 }} & {\footnotesize{0.05001 }} &  &  & \tabularnewline
{\footnotesize{60 }} & {\footnotesize{4 }} & {\footnotesize{1,4}} & {\footnotesize{3.162908 }} & {\footnotesize{0.04922 }} & {\footnotesize{3.162907 }} & {\footnotesize{0.05094 }} &  &  & \tabularnewline
{\footnotesize{60 }} & {\footnotesize{6 }} & {\footnotesize{1,4}} & {\footnotesize{3.051470 }} & {\footnotesize{0.04953 }} & {\footnotesize{3.051469 }} & {\footnotesize{0.05101 }} &  &  & \tabularnewline
{\footnotesize{80 }} & {\footnotesize{8 }} & {\footnotesize{1,4}} & {\footnotesize{3.090356 }} & {\footnotesize{0.04906 }} & {\footnotesize{3.090355 }} & {\footnotesize{0.05023 }} &  &  & \tabularnewline
{\footnotesize{800 }} & {\footnotesize{3 }} & {\footnotesize{1,4}} & {\footnotesize{3.887512 }} & {\footnotesize{0.04992 }} & {\footnotesize{3.887511 }} & {\footnotesize{0.05091 }} &  &  & \tabularnewline
{\footnotesize{1000 }} & {\footnotesize{1 }} & {\footnotesize{1,4}} & {\footnotesize{4.047191 }} & {\footnotesize{0.04944 }} & {\footnotesize{4.047190 }} & {\footnotesize{0.05094 }} &  &  & \tabularnewline
{\footnotesize{1000 }} & {\footnotesize{8 }} & {\footnotesize{1,4}} & {\footnotesize{3.814122 }} & {\footnotesize{0.04944 }} & {\footnotesize{3.814121 }} & {\footnotesize{0.05002 }} &  &  & \tabularnewline
\hline 
\end{tabular}
\par\end{centering}{\footnotesize \par}

\centering{}{\footnotesize{\caption{{\footnotesize{Critical values and exact alpha levels for those combinations
of $T$, $D$ and $M$ for which there does not exist a critical value
for $\alpha=0.05$. $T$ is the upper limit on the sample size (length
of surveillance), expressed in terms of the expected number of events
under the null. $D$ is the sample size at which the sequential analyses
start, also expressed in terms of the expected number of events under
the null. $M$ is the minimum number of events required to signal.
$CV_{cons}$ and $CV_{lib}$ are the conservative and liberal critical
values, respectively, while $\alpha_{cons}$ and $\alpha_{lib}$ are
their corresponding alpha levels.\label{CVdelayNC} }}}
}}
\end{table}

With a delayed start, there are some values of $T$ and $D$ for which
there is no critical value that gives an alpha level of exactly 0.05.
For those combinations, denoted with italics, Table~\ref{CVdelay}
presents the critical value that gives the largest possible alpha
less than 0.05. In Table~\ref{CVdelayNC}, we present the exact alpha
levels obtained for those scenarios, as well as the $\alpha>0.05$
obtained for a slightly smaller liberal critical value.

The exact critical values are based on numerical calculations done
in the same iterative way as for the original MaxSPRT
and the version described in the previous section. The only difference
is that there is an added initial step where the probabilities are
calculated for different number of events at the defined start time
$D$. Open source R functions\cite{R} have been published as part of the R package 
'Sequential' (cran.r-project.org/web/packages/Sequential/).

\subsection{Statistical Power and Timeliness}

\begin{table}
\centering{}{\footnotesize{}}%
\begin{tabular}{cccccccccrrrrr}
\hline 
 &  &  & \multicolumn{5}{c}{{\footnotesize{Power}}} &  & \multicolumn{5}{c}{{\footnotesize{Average Time to Signal}}}\tabularnewline
\cline{3-14} 
\multicolumn{2}{c}{{\footnotesize{RR=}}} &  & {\footnotesize{1.5 }} & {\footnotesize{2 }} & {\footnotesize{3 }} & {\footnotesize{4 }} & {\footnotesize{10 }} &  & {\footnotesize{1.5 }} & {\footnotesize{2 }} & {\footnotesize{3 }} & {\footnotesize{4 }} & {\footnotesize{10 }}\tabularnewline
{\footnotesize{T }} & {\footnotesize{D }} &  &  &  &  &  &  &  &  &  &  &  & \tabularnewline
{\footnotesize{5 }} & {\footnotesize{0 }} & {\footnotesize{\vline }} & {\footnotesize{0.190 }} & {\footnotesize{0.447 }} & {\footnotesize{0.876 }} & {\footnotesize{0.987 }} & {\footnotesize{1.000 }} &  & {\footnotesize{1.82 }} & {\footnotesize{2.09 }} & {\footnotesize{1.78 }} & {\footnotesize{1.22 }} & {\footnotesize{0.26}}\tabularnewline
{\footnotesize{5 }} & {\footnotesize{3 }} & {\footnotesize{\vline }} & {\footnotesize{0.275 }} & {\footnotesize{0.595 }} & {\footnotesize{0.943 }} & {\footnotesize{0.996 }} & {\footnotesize{1.000 }} &  & {\footnotesize{3.81 }} & {\footnotesize{3.65 }} & {\footnotesize{3.30 }} & {\footnotesize{3.08 }} & {\footnotesize{3.00 }}\tabularnewline
{\footnotesize{10 }} & {\footnotesize{0 }} & {\footnotesize{\vline}} & {\footnotesize{0.280 }} & {\footnotesize{0.685 }} & {\footnotesize{0.989 }} & {\footnotesize{1.000 }} & {\footnotesize{1.000 }} &  & {\footnotesize{4.02 }} & {\footnotesize{4.13 }} & {\footnotesize{2.45 }} & {\footnotesize{1.35 }} & {\footnotesize{0.27 }}\tabularnewline
{\footnotesize{10 }} & {\footnotesize{3 }} & {\footnotesize{\vline }} & {\footnotesize{0.377 }} & {\footnotesize{0.789 }} & {\footnotesize{0.996 }} & {\footnotesize{1.000 }} & {\footnotesize{1.000 }} &  & {\footnotesize{5.33 }} & {\footnotesize{4.84 }} & {\footnotesize{3.53 }} & {\footnotesize{3.10 }} & {\footnotesize{3.00 }}\tabularnewline
{\footnotesize{10 }} & {\footnotesize{6 }} & {\footnotesize{\vline }} & {\footnotesize{0.408 }} & {\footnotesize{0.819 }} & {\footnotesize{0.997 }} & {\footnotesize{1.000 }} & {\footnotesize{1.000 }} &  & {\footnotesize{6.94 }} & {\footnotesize{6.59 }} & {\footnotesize{6.07 }} & {\footnotesize{6.00 }} & {\footnotesize{6.00 }}\tabularnewline
{\footnotesize{20 }} & {\footnotesize{0 }} & {\footnotesize{\vline }} & {\footnotesize{0.450 }} & {\footnotesize{0.921 }} & {\footnotesize{1.000 }} & {\footnotesize{1.000 }} & {\footnotesize{1.000 }} &  & {\footnotesize{8.68 }} & {\footnotesize{6.96 }} & {\footnotesize{2.67 }} & {\footnotesize{1.41 }} & {\footnotesize{0.28 }}\tabularnewline
{\footnotesize{20 }} & {\footnotesize{3 }} & {\footnotesize{\vline }} & {\footnotesize{0.543 }} & {\footnotesize{0.952 }} & {\footnotesize{1.000 }} & {\footnotesize{1.000 }} & {\footnotesize{1.000 }} &  & {\footnotesize{9.44 }} & {\footnotesize{7.06 }} & {\footnotesize{3.78 }} & {\footnotesize{3.17 }} & {\footnotesize{3.00 }}\tabularnewline
{\footnotesize{20 }} & {\footnotesize{6 }} & {\footnotesize{\vline}} & {\footnotesize{0.583 }} & {\footnotesize{0.963 }} & {\footnotesize{1.000 }} & {\footnotesize{1.000 }} & {\footnotesize{1.000 }} &  & {\footnotesize{10.42 }} & {\footnotesize{8.20 }} & {\footnotesize{6.15 }} & {\footnotesize{6.01 }} & {\footnotesize{6.00 }}\tabularnewline
{\footnotesize{20 }} & {\footnotesize{10 }} & {\footnotesize{\vline }} & {\footnotesize{0.609 }} & {\footnotesize{0.969 }} & {\footnotesize{1.000 }} & {\footnotesize{1.000 }} & {\footnotesize{1.000 }} &  & {\footnotesize{12.33 }} & {\footnotesize{10.83 }} & {\footnotesize{10.01 }} & {\footnotesize{10.00 }} & {\footnotesize{10.00 }}\tabularnewline
{\footnotesize{50 }} & {\footnotesize{0 }} & {\footnotesize{\vline }} & {\footnotesize{0.803 }} & {\footnotesize{1.000 }} & {\footnotesize{1.000 }} & {\footnotesize{1.000 }} & {\footnotesize{1.000 }} &  & {\footnotesize{20.45 }} & {\footnotesize{8.94 }} & {\footnotesize{2.82 }} & {\footnotesize{1.48 }} & {\footnotesize{0.30 }}\tabularnewline
{\footnotesize{50 }} & {\footnotesize{3 }} & {\footnotesize{\vline }} & {\footnotesize{0.860 }} & {\footnotesize{1.000 }} & {\footnotesize{1.000 }} & {\footnotesize{1.000 }} & {\footnotesize{1.000 }} &  & {\footnotesize{19.39 }} & {\footnotesize{8.50 }} & {\footnotesize{3.85 }} & {\footnotesize{3.18 }} & {\footnotesize{3.00 }}\tabularnewline
{\footnotesize{50 }} & {\footnotesize{6 }} & {\footnotesize{\vline }} & {\footnotesize{0.871 }} & {\footnotesize{1.000 }} & {\footnotesize{1.000 }} & {\footnotesize{1.000 }} & {\footnotesize{1.000 }} &  & {\footnotesize{19.65 }} & {\footnotesize{9.43 }} & {\footnotesize{6.16 }} & {\footnotesize{6.01 }} & {\footnotesize{6.00 }}\tabularnewline
{\footnotesize{50 }} & {\footnotesize{10 }} & {\footnotesize{\vline }} & {\footnotesize{0.885 }} & {\footnotesize{1.000 }} & {\footnotesize{1.000 }} & {\footnotesize{1.000 }} & {\footnotesize{1.000 }} &  & {\footnotesize{20.64 }} & {\footnotesize{11.82 }} & {\footnotesize{10.02 }} & {\footnotesize{10.00 }} & {\footnotesize{10.00 }}\tabularnewline
{\footnotesize{100 }} & {\footnotesize{0 }} & {\footnotesize{\vline }} & {\footnotesize{0.978 }} & {\footnotesize{1.000 }} & {\footnotesize{1.000 }} & {\footnotesize{1.000 }} & {\footnotesize{1.000 }} &  & {\footnotesize{29.93 }} & {\footnotesize{9.30 }} & {\footnotesize{2.92 }} & {\footnotesize{1.53 }} & {\footnotesize{0.31 }}\tabularnewline
{\footnotesize{100 }} & {\footnotesize{3 }} & {\footnotesize{\vline }} & {\footnotesize{0.987 }} & {\footnotesize{1.000 }} & {\footnotesize{1.000 }} & {\footnotesize{1.000 }} & {\footnotesize{1.000 }} &  & {\footnotesize{27.16 }} & {\footnotesize{8.95 }} & {\footnotesize{3.90 }} & {\footnotesize{3.18 }} & {\footnotesize{3.00 }}\tabularnewline
{\footnotesize{100 }} & {\footnotesize{6 }} & {\footnotesize{\vline}} & {\footnotesize{0.988 }} & {\footnotesize{1.000 }} & {\footnotesize{1.000 }} & {\footnotesize{1.000 }} & {\footnotesize{1.000 }} &  & {\footnotesize{26.98 }} & {\footnotesize{9.97 }} & {\footnotesize{6.24 }} & {\footnotesize{6.01 }} & {\footnotesize{6.00 }}\tabularnewline
{\footnotesize{100 }} & {\footnotesize{10 }} & {\footnotesize{\vline }} & {\footnotesize{0.990 }} & {\footnotesize{1.000 }} & {\footnotesize{1.000 }} & {\footnotesize{1.000 }} & {\footnotesize{1.000 }} &  & {\footnotesize{27.40 }} & {\footnotesize{12.09 }} & {\footnotesize{10.02 }} & {\footnotesize{10.00 }} & {\footnotesize{10.00 }}\tabularnewline
{\footnotesize{200 }} & {\footnotesize{0 }} & {\footnotesize{\vline}} & {\footnotesize{1.000 }} & {\footnotesize{1.000 }} & {\footnotesize{1.000 }} & {\footnotesize{1.000 }} & {\footnotesize{1.000 }} &  & {\footnotesize{33.00 }} & {\footnotesize{9.62 }} & {\footnotesize{3.01 }} & {\footnotesize{1.58 }} & {\footnotesize{0.32 }}\tabularnewline
{\footnotesize{200 }} & {\footnotesize{3 }} & {\footnotesize{\vline }} & {\footnotesize{1.000 }} & {\footnotesize{1.000 }} & {\footnotesize{1.000 }} & {\footnotesize{1.000 }} & {\footnotesize{1.000 }} &  & {\footnotesize{30.01 }} & {\footnotesize{9.35 }} & {\footnotesize{3.94 }} & {\footnotesize{3.18 }} & {\footnotesize{3.00 }}\tabularnewline
{\footnotesize{200 }} & {\footnotesize{6 }} & {\footnotesize{\vline }} & {\footnotesize{1.000 }} & {\footnotesize{1.000 }} & {\footnotesize{1.000 }} & {\footnotesize{1.000 }} & {\footnotesize{1.000 }} &  & {\footnotesize{29.78 }} & {\footnotesize{10.31 }} & {\footnotesize{6.26 }} & {\footnotesize{6.01 }} & {\footnotesize{6.00 }}\tabularnewline
{\footnotesize{200 }} & {\footnotesize{10 }} & {\footnotesize{\vline }} & {\footnotesize{1.000 }} & {\footnotesize{1.000 }} & {\footnotesize{1.000 }} & {\footnotesize{1.000 }} & {\footnotesize{1.000 }} &  & {\footnotesize{30.16 }} & {\footnotesize{12.48 }} & {\footnotesize{10.04 }} & {\footnotesize{10.00 }} & {\footnotesize{10.00 }}\tabularnewline
\hline 
\end{tabular}{\footnotesize{\caption{{\footnotesize{Statistical power and average time to signal for the
Poisson based maximized SPRT, when the analyses does not start until
the sample size is large enough to correspond to $D$ expected events
under the null hypothesis. $T$ is the upper limit on the sample size
(length of surveillance), expressed in terms of the expected number
of events under the null. The minimum number of events required to
signal is set to $M=1$. The type 1 error is $\alpha=0.05$. \label{PTdelay} }}}
}}
\end{table}

For a fixed value on the upper limit on the sample size $T$, the
statistical power of sequential analyses always increases if there
are fewer looks at the data, with the maximum attained when there
is only one non-sequential analysis after all the data has been collected.
Hence, for fixed $T$, a delay in the start of surveillance always
increases power, as can be seen in Table~\ref{PTdelay}. For fixed
$T$, the average time to signal almost always increases with a delayed
start. The rare exception is when $T$ is very large and the true
RR is very small. For example, for $T=100$ and $RR=1.5$, the average
time to signal is 29.9 without a delayed start, 27.2 with a delayed
start of $D=3$ and 27.0 with a delayed start of $D=6$. With a longer
delay of $D=10$, the average time to signal increases to 27.4.

\begin{figure}[tbp] 
  \centering
  \includegraphics[bb=0 0 1280 923,width=5.67in,height=4.09in,keepaspectratio]{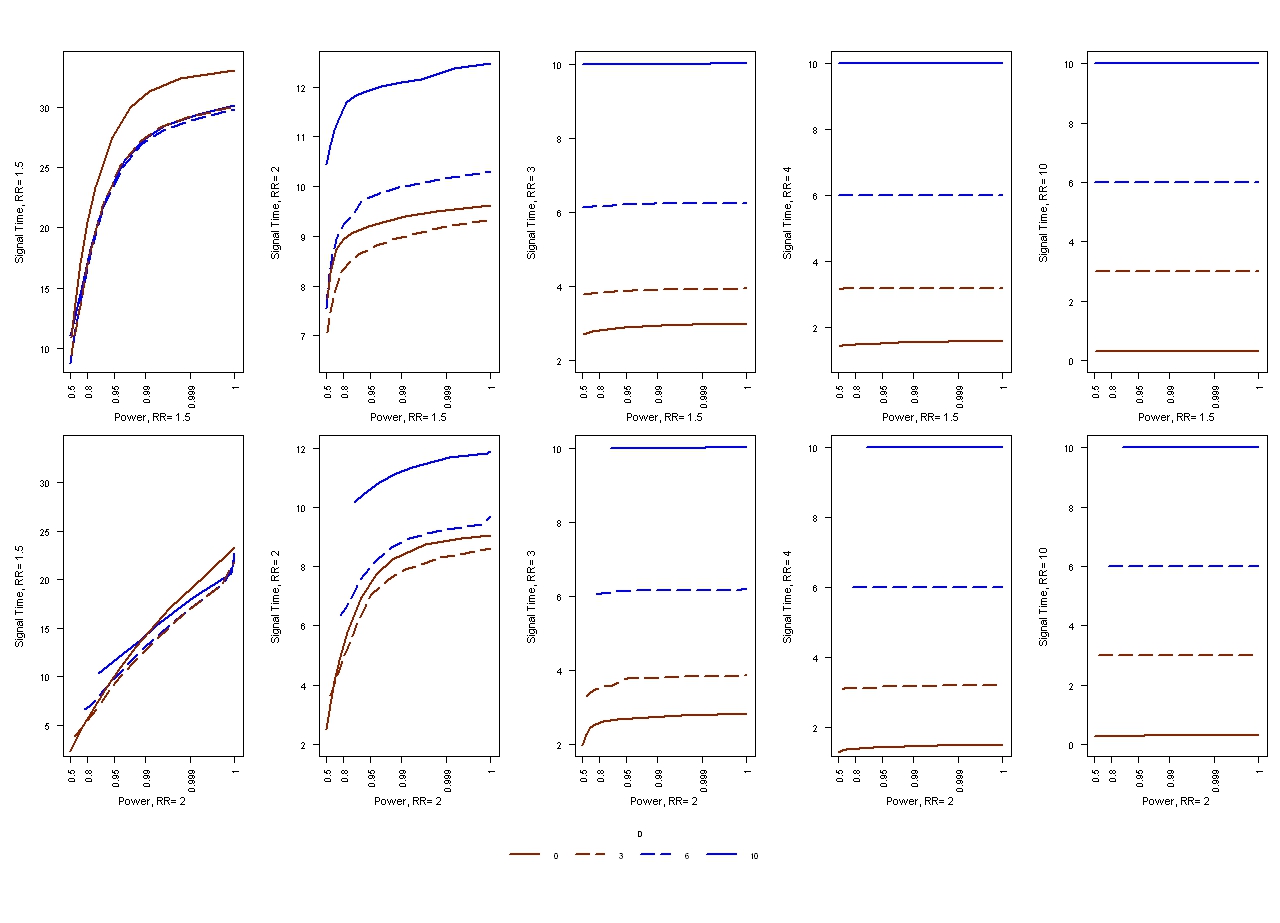}
  \caption{The average time to signal, as a function of statistical power, for
the Poisson based maximized SPRT, when the analyses does not start
until the sample size is large enough to correspond to $D$ expected
events under the null hypothesis. The type 1 error is $\alpha=0.05$.
\label{delay}}
\end{figure}

For fixed $T$, we saw that there is a trade-off between power and
the time to signal, but in post-market safety surveillance it is usually
easy and inexpensive to increase power by increasing $T$. Hence,
the critical evaluation is to compare the average time to signal when
holding both power and the alpha level fixed. This is done in Figure~\ref{delay}.
When the study is powered for a relative risk of 2, then the average
time to signal is lower when there is less of a delay in the start
of the surveillance, whether the true relative risk is small or large.
When the study is powered for a relative risk of 1.5, we see the same
thing, except when the true relative risk is small. Hence, in terms
of performance, smaller $D$ is always better.

\section{Discussion}

With the establishment of new near real-time post-market drug and safety surveillance systems\cite{Fireman-PDS2012, Burwen-AJPH2012, Hauser-CCQO2012, Huang-Vaccine2010, Nguyen-PDS2012}, 
sequential statistical methods will become a standard feature of the phramacovigilance landscape.
In this paper we have shown that it is possible to reduce the expected time to signal when the
null is rejected, without loss of statistical power, by requiring a minimum number of adverse events 
before generating a statistical signal. This will allow users to optimize their post-market sequential analyses. 

In this paper we calculated the critical values, power and timeliness
for Poisson based continuous sequential analysis with either a minimum events to signal 
requirement or when there is delayed start for logistical reasons.
The reported numbers are based on exact numerical calculations rather
than approximate asymptotic calculations or computer simulations.
From a mathematical and statistical perspective, these are straight
forward extensions of prior work on exact continuous sequential analysis.
The importance of the results are hence from practical public health perspective 
rather than for any theoretical statistical advancements. 

A key question is which sequential study design to use. There is not
always a simple answer to that question, as the performance of the
various versions depends on the true relative risk, which is unknown.
One important consideration is that the early detection of an adverse
event problem is more important when the relative risk is high, since
more patients are affected. As a rule of thumb, it is reasonable to
require a minimum of about $M=3$ to $6$ adverse events before rejecting
the null hypothesis, irrespectively of whether it is a rare or common
adverse event. For those who want a specific recommendation, we suggest
$M=4$. 

Critical values, statistical power
and average time to signal has been presented for a wide variety of parameter values.
This is done so that most user will not have to perform their own
calculations. For those who want to use other parameter values,
critical values, power and expected time to signal can be calculated 
using the 'Sequential' R package that we have developed. 

It is possible to combine a delayed start with $D>0$ together with
a requirement that there are at least $M>1$ events to signal. It
does not always make a difference though. For $M=4$, the critical
values are the same as for $M=1$, for all values of $D\geq1$. That
is because with $D=1$ or higher, one would never signal with less
than three events anyhow. Since the critical values are the same,
the statistical power and average time to signal are also the same.
This means that when there is a non-trivial delayed start, there
is not much benefit from also requiring a minimum number events to
signal, but the 'Sequential' R package has a function for this dual scenario as well.

There is no reason to purposely delay the start of the surveillance
until there is some minimum sample size $D$. In the few scenarios
for which such a delay improve the performance, the improvement is
not measurably better than the improvements obtained by using a minimum
number of observed events. Only when it is logistically impossible
to start the surveillance at the very beginning should such sequential
analyses be conducted, and then it is important to do so in order
to maximize power, to minimize the time to signal and to maintain
the correct alpha level.

For self-controlled analyses, a binomial version of the MaxSPRT\cite{ku-SA2011} is
used rather than the Poisson version discussed in this paper.
For concurrent matched controls, a flexible exact sequential method
is used that allows for a different number of controls per exposed
individuals\cite{Fireman-M2013}. By default, these types of continuous
sequential methods will not reject the null hypothesis until there
is a minimum number of events observed. To see this, consider the
case with a 1:1 ratio of exposed to unexposed and and assume that
the first four adverse events all are in the exposed category. Under
the null hypothesis, the probability of this is $(1/2)^{4}=0.0625$,
which does not give a low enough p-value to reject the null hypothesis
even in a non-sequential setting. Hence, the null will never be rejected
after only four adverse events, even when there is no minimum requirement.
One could set the minimum number of exposed events to something higher,
and that may be advantageous. If there is a delayed start for logistical reasons, then it
makes sense to take that into account when calculating the critical
value, for these two types of models as well.

Since the Vaccine Safety Datalink\cite{Chen-Pediatrics1997} launched
the first near real-time post-marketing vaccine safety surveillance
system in 2004\cite{Davis-Pharmaceutics2013}, continuous sequential analysis has been used for a
number of vaccines and potential adverse events\cite{Yih-Pediatrics2011, Lieu-MC2007,Yih-Vaccine2009,Belongia-PIDJ2010,Klein-Pediatrics2010,Gee-Vaccine2011,Lee-AJPM2011, Weintraub-NEJM2014}.
The critical value tables presented in this paper has already been
used by the Vaccine Safety Datalink project. As new near real-time post-market
safety surveillance systems are being developed,  it is important to fine-tune and optimize
the performance of near-real time safety surveillance 
systems\cite{Fireman-PDS2012, Suling-Pharmaceutics2012, Hauser-CCQO2012, Gaalen-PDS2014, Li-SM2014, Maro-PDS2014}. 
While the improved time to signal is modest compared to the original
version of the Poisson based MaxSPRT, there is no reason not to use
these better designs.

\section*{Acknowledgments}

This work was supported by the Centers for Disease Control and Prevention
through the Vaccine Safety Datalink Project, contract number 200-2002-00732
(M.K.), by National Institute of General Medical Sciences grant 1R01GM108999 (M.K.), 
by the Conselho Nacional de Desenvolvimento Científico e Tecnológico, Brazil (I.S.), 
and by the Banco de Desenvolvimento de Minas Gerais, Brazil (I.S.).

\global\long\def\baselinestretch{1.0}


\begin{thebibliography}{10}


\bibitem{Yih-Pediatrics2011}
Yih WK, Kulldorff M, Fireman BH, Shui IM, Lewis EM, Klein NP, Baggs J,
  Weintraub ES, Belongia EA, Naleway A, Gee J, Platt R, and Lieu TA.
\newblock Active surveillance for adverse events: The experience of the vaccine
  safety datalink project.
\newblock {\em Pediatrics}, 127:S54--64, 2011.

\bibitem{Davis-Pharmaceutics2013}
Davis RL.
\newblock  Vaccine Safety Surveillance Systems: Critical Elements and Lessons Learned in the Development of the US Vaccine Safety Datalink's Rapid Cycle Analysis Capabilities.
\newblock {\em Pharmaceutics}, 5:168--178, 2013.

\bibitem{McNeil-Vaccine2014}
McNeil MM, Gee J, Weintraub ES, Belongia EA, Lee GM, Glanz JM, Nordin JD, Klein NP, Baxter R, Naleway AL, Jackson LA, Omer SB, Jacobsen SJ, DeStefano F. 
\newblock The Vaccine Safety Datalink: successes and challenges monitoring vaccine safety. 
\newblock {\em Vaccine}, 32:5390--5398, 2014.

\bibitem{ku-SA2011}
Kulldorff M., Davis RL, Kolczak M, Lewis E, Lieu T, and Platt R.
\newblock A maximized sequential probability ratio test for drug and vaccine
  safety surveillance.
\newblock {\em SA}, 3000:58--78, 2011.

\bibitem{Lieu-MC2007}
Lieu TA, Kulldorff M, Davis RL, Lewis EM, Weintraub E, Yih WK, Yin R, Brown JS,
  and Platt R.
\newblock Real-time vaccine safety surveillance for the early detection of
  adverse events.
\newblock {\em Medical Care}, 45:S89--95, 2007.

\bibitem{Yih-Vaccine2009}
Yih WK, Nordin JD, Kulldorff M, Lewis E, Lieu T, Shi P, and Weintraub E.
\newblock An assessment of the safety of adolescent and adult
  tetanus-diphtheria-acellular pertussis (tdap) vaccine, using near real-time
  surveillance for adverse events in the vaccine safety datalink.
\newblock {\em Vaccine}, 27:4257--4262, 2009.

\bibitem{Belongia-PIDJ2010}
Belongia EA, Irving SA, Shui IM, Kulldorff M, Lewis E, Li~R, Lieu TA, Weintraub
  E, Yih WK, Yin R, Baggs J, and the Vaccine Safety Datalink
  Investigation~Group.
\newblock Real-time surveillance to assess risk of intussusception and other
  adverse events after pentavalent, bovine-derived rotavirus vaccine.
\newblock {\em Pediatric Infectious Disease Journal}, 29:1--5, 2010.

\bibitem{Klein-Pediatrics2010}
Klein NP, Fireman B, Yih WK, Lewis E, Kulldorff M, Ray P, Baxter R, Hambidge S,
  Nordin J, Naleway A, Belongia EA, Lieu T, Baggs J, Weintraub E, and for~the
  Vaccine Safety~Datalink.
\newblock Measles-mumps-rubella-varicella combination vaccine and the risk of
  febrile seizures.
\newblock {\em Pediatrics}, 126:e1--8, 2010.

\bibitem{Gee-Vaccine2011}
Gee J, Naleway A, Shui I, Baggs J, Yin R, Li R, Kulldorff M, Lewis E, Fireman B, Daley MF, Klein NP, Weintraub ES. (2011). \newblock Monitoring the safety of quadrivalent human papillomavirus vaccine: findings from the Vaccine Safety Datalink. 
\newblock {\em Vaccine}, 29:8279--8284, 2011.

\bibitem{Lee-AJPM2011}
Lee GM, Greene SK, Weintraub ES, Baggs J, Kulldorff M, Fireman BH, Baxter R, Jacobsen SJ, Irving S, Daley MF, Yin R, Naleway A, Nordin J, Li L, McCarthy N, Vellozi C, DeStefano F, Lieu TA, on behalf of the Vaccine Safety Datalink Project. 
\newblock H1N1 and Seasonal Influenza in the Vaccine Safety Datalink Project. 
\newblock {\em American Journal of Preventive Medicine}, 41:121--128, 2011.

\bibitem{Tseng-Vaccine2013}
Tseng HF, Sy LS, Liu ILA, Qian L, Marcy SM, Weintraub E, Yih K, Baxter R, Glanz J, Donahue J, Naleway A, Nordin J, Jacobsen SJ. 
\newblock  Postlicensure surveillance for pre-specified adverse events following the 13-valent pneumococcal conjugate vaccine in children. 
\newblock {\em Vaccine}, 31, 2578--2583, 2013.

\bibitem{Weintraub-NEJM2014}
Weintraub ES, Baggs J, Duffy J, Vellozzi C, Belongia EA, Irving S, Klein NP, Glanz J, Jacobsen SJ, Naleway A, Jackson LA, DeStefano F. (2014). 
\newblock Risk of intussusception after monovalent rotavirus vaccination. 
\newblock {\em New England Journal of Medicine}, 370:513--519, 2014.

\bibitem{Daley-Vaccine2014}
Daley MF, Yih WK, Glanz JM, Hambidge SJ, Narwaney KJ, Yin R, Li L, Nelson JC, Nordin JD, Klein NP, Jacobsen SJ, Weintraub E. 
\newblock Safety of diphtheria, tetanus, acellular pertussis and inactivated poliovirus (DTaP-IPV) vaccine. 
\newblock {\em Vaccine}, 32:3019--3024, 2014.

\bibitem{Brown-PDS2007}
Brown JS, Kulldorff M, Chan KA, Davis RL, Graham D, Pettus PT, Andrade SE,
  Raebel M, Herrinton L, Roblin D, Boudreau D, Smith D, Gurwitz JH, Gunter MJ,
  and Platt R.
\newblock Early detection of adverse drug events within population-based health
  networks: Application of sequential testing methods.
\newblock {\em Pharmacoepidemiology and Drug Safety}, 16:1275--1284, 2007.

\bibitem{Avery-PDS2013}
Avery T, Kulldorff M, Vilk W, Li~L, Cheetham C, Dublin S, Hsu J, Davis RL, Liu
  L, Herrinton L, Platt R, and Brown JS.
\newblock Near real-time adverse drug reaction surveillance within
  population-based health networks: methodology considerations for data
  accrual.
\newblock {\em Pharmacoepidemiology and Drug Safety}, page epub, 2013.

\bibitem{Fireman-PDS2012}
Fireman B, Toh S, Butler MG, Go AS, Joffe HV, Graham DJ, Nelson JC, Daniel GW, Selby JV. 
\newblock A protocol for active surveillance of acute myocardial infarction in association with the use of a new antidiabetic pharmaceutical agent. 
\newblock {\em Pharmacoepidemiology and drug safety}, 21, 282--290, 2012.

\bibitem{Suling-Pharmaceutics2012}
Suling M, Pigeot I.
\newblock Signal Detection and Monitoring Based on Longitudinal Healthcare Data
\newblock {\em Pharmaceutics}, 4:607--640, 2012.

\bibitem{Gagne-Epidemiology2012}
Gagne JJ, Rassen JA, Walker AM, Glynn RJ, Schneeweiss S. 
\newblock Active safety monitoring of new medical products using electronic healthcare data: selecting alerting rules.
\newblock {\em Epidemiology}, 23:238--246, 2012.

\bibitem{Gagne-PDS2014}
Gagne JJ, Wang SV, Rassen JA, Schneeweiss S. 
\newblock A modular, prospective, semi-automated drug safety monitoring system for use in a distributed data environment. 
\newblock {\em Pharmacoepidemiology and Drug Safety}, 23:619--627, 2014.

\bibitem{Wald-AMS1945}
Wald A.
\newblock Sequential tests of statistical hypotheses.
\newblock {\em Annals of Mathematical Statistics}, 16:117--186, 1945.

\bibitem{Wald-book1947}
Wald A.
\newblock {\em Sequential Analysis}.
\newblock Wiley, 1947.

\bibitem{Weiss-AMS1953}
Weiss L.
\newblock Testing one simple hypothesis against another.
\newblock {\em Annals of Mathematical Statistics}, 24:273--281, 1953.

\bibitem{Lai-Ghosh1991}
Lai TL.
\newblock Asymptotic optimality of generalized sequential likelihood ratio tests in some classical sequential testing procedures.
\newblock Ghosh BK and Sen PK (eds), Handbook of Sequential Analysis, 121--144.
\newblock {\em Dekker: New York}, 1991.

\bibitem{Siegmund-AS1980}
Siegmund D and Gregory P.
\newblock A sequential clinical trial for testing p1 = p2.
\newblock {\em Annals of Statistics}, 8:1219--1228, 1980.

\bibitem{R}
{R Development Core Team}.
\newblock {\em R: A Language and Environment for Statistical Computing}.
\newblock R Foundation for Statistical Computing, Vienna, Austria, 2009.
\newblock {ISBN} 3-900051-07-0.

\bibitem{Burwen-AJPH2012}
Burwen DR, Sandhu SK, MaCurdy TE, Kelman JA, Gibbs JM, Garcia B, Marakatou M, Forshee RA, Izurieta HS, Ball R.
\newblock  Surveillance for Guillain-Barre syndrome after influenza vaccination among the Medicare population, 2009-2010.
\newblock {\em American Journal of Public Health}, 102, 1921--1927, 2012.

\bibitem{Hauser-CCQO2012}
Hauser RG, Mugglin AS, Friedman PA, Kramer DB, Kallinen L, McGriff D, Hayes DL. 
\newblock  Early detection of an underperforming implantable cardiovascular device using an automated safety surveillance tool. 
\newblock {\em Circulation: Cardiovascular Quality and Outcomes},  5:189--196, 2012.

\bibitem{Huang-Vaccine2010}
Huang WT, Chen WW, Yang HW, Chen WC, Chao YN, Huang YW, Chuang JH, Kuo HS.
\newblock Design of a robust infrastructure to monitor the safety of the pandemic A (H1N1) 2009 vaccination program in Taiwan.
\newblock {\em Vaccine}, 28, 7161--7166, 2010.

\bibitem{Nguyen-PDS2012}
Nguyen M, Ball R, Midthun K, Lieu TA. 
\newblock The Food and Drug Administration's Post-Licensure Rapid Immunization Safety Monitoring program: strengthening the federal vaccine safety enterprise. 
\newblock {\em Pharmacoepidemiology and Drug Safety}, 21:291--297, 2012.

\bibitem{Fireman-M2013}
Fireman B, et al.
\newblock Exact sequential analysis for binomial data with time-varying probabilities.
\newblock {\em Manuscript in Preparation}, 2013.

\bibitem{Chen-Pediatrics1997}
Chen RT, JE~Glasser JE, PH~Rhodes PH, RL~Davis RL, WE~Barlow WE, RS~Thompson, JP~Mullooloy,
  SB~Black, HR~Shinefield, CM~Vadheim, SM~Marcy, JI~Ward, RP~Wise, SG~Wassilak,
  SC~Hadler, and the Vaccine Safety Datalink~Team.
\newblock Vaccine safety datalink project: A new tool for improving vaccine
  safety monitoring in the united states.
\newblock {\em Pediatrics}, 99:765--773, 1997.

\bibitem{Gaalen-PDS2014}
Gaalen RD, Abrahamowicz M, Buckeridge DL. 
\newblock The impact of exposure model misspecification on signal detection in prospective pharmacovigilance. 
\newblock {\em Pharmacoepidemiology and Drug Safety}, 2014.

\bibitem{Li-SM2014}
Li R, Stewart B, Weintraub E, McNeil MM. 
\newblock Continuous sequential boundaries for vaccine safety surveillance. 
\newblock {\em Statistics in Medicine}, 33:3387--3397, 2014.

\bibitem{Maro-PDS2014}
Maro JC, Brown JS, Dal Pan GJ, Kulldorff M. 
\newblock Minimizing signal detection time in postmarket sequential analysis: balancing positive predictive value and sensitivity. 
\newblock {\em Pharmacoepidemiology and Drug Safety}, 23:839--848, 2014.


\end{thebibliography}
\end{document}